\documentstyle[preprint,aps]{revtex}
\begin{document}
\draft
\preprint{ }
\title{Linear-Response Calculation of Electron-Phonon
Coupling Parameters}
\author{Amy Y. Liu }
\address{ Department of Physics, Georgetown University,
Washington, DC 20057-0995}
\author{Andrew A. Quong}
\address{Sandia National Laboratories, Livermore, CA 94551-0969}
\maketitle
\begin{abstract}

An {\it ab initio}  method for calculating electron-phonon coupling
parameters is presented.   The method is an extension of the
plane-wave-based linear-response method for the calculation of
lattice dynamics.   Results for the mass enhancement parameter
$\lambda$ and the electron-phonon spectral function
$\alpha^2F(\omega)$ for  Al, Pb and Li are presented.
Comparisons are made to available experimental data.

\end{abstract}


\narrowtext
\section{Introduction}

The electron-phonon interaction in metals
plays an important role in a
variety of experimentally accessible quantities including
the enhancement of the electron mass,
the phonon lifetime arising from electron-phonon scattering,
electrical and thermal conductivities, and  the
superconducting transition temperature.
The  electron-phonon spectral function $\alpha^2F(\omega)$
measures the effectiveness of phonons with energy $\omega$ to
scatter electrons from one part of the Fermi surface to
another part.  Once it and the Coulomb pseudopotential
$\mu^*$ are determined for a superconducting material, all
of the thermodynamic properties of the superconductor,
including the superconducting gap as a function of temperature,
the transition temperature, and the discontinuity in the specific
heat at $T_c$, can be computed \cite{Carbotte90}.
In addition, transport properties of materials in the
normal state can be calculated from the
closely related spectral function $\alpha_{tr}^2F(\omega)$.

The ability to accurately  calculate  electron-phonon
coupling parameters has long  been a sought after goal.
It is a formidable task requiring
knowledge of the  the low-energy electronic excitation
spectrum, the complete vibrational spectrum, and the
self-consistent response of the electronic system to lattice
vibrations. {\it Ab initio} calculations of electron-phonon
coupling parameters  have proceeded along two distinct lines.
In the rigid ion \cite{RI} (RI) and
rigid muffin tin \cite{RMT} (RMT) schemes, the
screened electron-phonon potential is approximated by
neglecting changes in the potential everywhere except within the
atomic sphere of the displaced atom.
While these non-self-consistent
approximations  appear to be adequate for many transition
metals \cite{Klein82}, their validity has been questioned in
some cases, especially for anisotropic or
low-density-of-states materials \cite{Winter81,Rainer79,Krakauer93}.
An alternative to the RI and RMT methods is the
frozen-phonon total-energy method \cite{Lam86}.
In this approach, the electron-phonon matrix
elements are evaluated using the self-consistently screened
potentials corresponding to frozen-in phonon displacements.
The primary drawback of the frozen-phonon approach is that
only phonon wavevectors that are commensurate
with the lattice and that correspond to reasonably sized supercells
can be considered.  This makes it difficult to determine accurately
quantities that involve integrations over the phonon wavevector
${\bf q}$ throughout the Brillouin zone.
These include, for example, the electron-phonon mass enhancement
parameter $\lambda$, the phonon density of states $F(\omega)$,  and
the electron-phonon spectral function $\alpha^2F(\omega)$.

Recently, linear-response
theory  within the framework of density-functional calculations
has been shown to be an efficient and powerful alternative
to the frozen-phonon method for
calculating  lattice dynamical properties of solids \cite{Baroni87}.
Atomic displacements are treated as
perturbations, and  the electronic response to
the perturbation is calculated self-consistently.
Perturbations of arbitrary wavevector  ${\bf q}$
can be treated  without using supercells.  The linear-response method
has been implemented with a variety of different basis
sets for representing the electronic wavefunctions, and it has been
successfully applied to the study of lattice dynamics in
a wide range of systems including
semiconductors \cite{Baroni87,Giannozzi91},
metals \cite{Quong92,Quong_mrs93,Metals},
ferroelectrics \cite{Krakauer95}, and surfaces \cite{Quong_surf}.

In this paper, we extend the plane-wave-based
density-functional linear-response method
to the calculation of  electron-phonon
coupling parameters.  A similar method based on linearized
muffin-tin orbitals was recently presented in Ref. 16.
The use of a plane-wave basis has the advantage of
simplicity both in terms of formalism and implementation.
We apply our method to the calculation of the
electron-phonon spectral function $\alpha^2F(\omega)$
for several elemental metals: Al, which is a well-studied
weak-coupling superconductor;  Pb, which is the prototypical
strong-coupling superconductor; and Li, which is not observed
to have a superconducting phase, but which
earlier calculations have found to have a
moderate electron-phonon mass enhancement parameter.

\section{Method}

For notational simplicity, we consider the case of a single atom of
mass $M$ per unit cell.
The electron-phonon matrix element for scattering of an
electron from a Bloch state $n{\bf k}$ to another
Bloch state $n^\prime{\bf k}^\prime$
by a phonon  of frequency $\omega_{{\bf k}-{\bf k}^\prime \nu}$
is
\begin{equation}
g(n{\bf k},n^\prime{\bf k}^\prime,\nu) =
 ({\hbar\over2M\omega_{{\bf k}-{\bf k}^\prime \nu}})^{1/2}
  \langle \psi_{n{\bf k}} |
  {\bf \hat{\epsilon}}_{{\bf  k}-{\bf k}^\prime \nu}
                \cdot{\bf \nabla_R} V_{sc}  |
 \psi_{n^\prime{\bf k}^\prime}\rangle ,
\end{equation}
where $ {\bf \hat{\epsilon}}_{{\bf k}-{\bf k}^\prime \nu}$
is the phonon polarization vector, and ${\bf \nabla_R} V_{sc}$ is
the gradient of the self-consistent potential with respect to
atomic displacements.
The linewidth of phonon ${\bf q}\nu$
arising from electron-phonon scattering is given by
\begin{equation}
\gamma_{{\bf q} \nu} = 2\pi\omega_{{\bf q}\nu}
\sum_{n,n^\prime,{\bf k},{\bf k}^\prime}
\delta(E_{n{\bf k}} - E_F) \delta(E_{n^\prime{\bf k}^\prime} - E_F)
\delta_{{\bf k}-{\bf k}^\prime-{\bf q}}
|g(n{\bf k},n^\prime{\bf k}^\prime,\nu)|^2 ,
\end{equation}
where $E_F$ is the Fermi energy.

For many applications, the quantity of interest involves a sum or
average of the electron-phonon coupling strength
over wavevectors throughout the Brillouin zone.    For example,
the electron-phonon spectral function $\alpha^2F(\omega)$, which
is a central quantity in the  strong-coupling
theory of superconductivity, is given by a sum over contributions
to the coupling from each phonon mode:
\begin{equation}
\alpha^2F(\omega) = {1\over 2\pi N(E_F) } \sum_{{\bf q}\nu}
\delta(\omega - \omega_{{\bf q}\nu})
{\gamma_{{\bf q}\nu} \over \hbar\omega_{{\bf q}\nu}}.
\end{equation}
Here $N(E_F)$ is the electronic density of states at the Fermi level.
The dimensionless electron-phonon mass enhancement parameter also
involves a sum over modes and can be expressed
as the first inverse frequency moment of the spectral function:
\begin{equation}
\lambda = 2\int d\omega \alpha^2F(\omega)/\omega.
\end{equation}

In this work,
the electronic wavefunctions  $\psi_{n{\bf k}}$  and eigenvalues
$E_{n{\bf k}}$ are calculated
using the {\it ab initio} pseudopotential local-density formalism.
The electron-ion interaction is represented by soft
separable pseudopotentials \cite{Troullier91}, and the single-particle
wavefunctions are expanded in a plane-wave basis set.
The Wigner form of the exchange and correlation
functional is employed \cite{Wigner}, and in the
case of Pb and Li, the partial core correction is used
to handle the nonlinearity of the exchange and correlation
interaction between the core and valence charge
densities \cite{partial_core}.
Unless otherwise indicated, the
calculations are performed at lattice constants  that are
determined within the local-density approximation (LDA) and that
are in good agreement with the experimental values.

The phonon frequencies and polarization vectors are calculated using
linear-response theory.  The second-order
change in the total energy, and hence the dynamical matrix, depends
only on the first-order change in the electronic charge density.
The linear response of the electronic density to atomic displacements
is determined self-consistently by solving a Bethe-Salpeter
equation as discussed in Ref. 11.  We have generalized the method
to include corrections for the overlap between core and valence
charge densities \cite{Quong_mrs93}.
The electron-phonon matrix elements,
$g(n{\bf k},n^\prime{\bf k}^\prime,\nu)$, are easily
computed from the first-order change in the self-consistent
potential.

The doubly-constrained  Fermi surface sums in Eq. (2)
are performed using dense meshes of 1300 and 728 ${\bf k}$ points
in the irreducible Brillouin zones (IBZ) of the fcc and bcc
structures, respectively.
The $\delta$ functions in energy are replaced
by Gaussians of width 0.02 Ry.   Because of the large number of
${\bf k}$ points sampled, the results are not very sensitive to
the Gaussian width.
Phonon wavevectors are sampled on coarser meshes of
89 and 140 points in the fcc and bcc IBZs, respectively.

\section{Results}

To test the accuracy of the method, we consider first the
simple metal Al.  The LDA gives very good structural properties
for fcc Al, and the linear-response method yields phonon
dispersion curves in excellent agreement with experiments throughout
the Brillouin zone \cite{Quong92}.
The $\alpha^2F(\omega)$ for Al calculated in this work
is shown in Fig. 1 (solid line), along with results from
two experiments (long- and short-dashed lines) \cite{Wolf}.
Extraction of $\alpha^2F$ from conventional
McMillan-Rowell tunneling spectroscopy is not possible
for Al since it is too close to an ideal BCS superconductor.
Instead the experimental spectral functions shown in
Fig. 1 were extracted from
proximity electron tunneling data \cite{Wolf}.
Unfortunately, the inversion of this type of tunneling data
involves the introduction of additional  fitting parameters
characterizing the proximity layer.  This introduces
uncertainties in the extracted spectral functions, as evidenced by
the differences between the two experimentally determined
$\alpha^2F$ functions shown in Fig. 1.
The calculated spectral function agrees
reasonably well with the experimental curves.

The value of the electron-phonon mass enhancement
parameter determined from the first inverse-frequency
moment of the calculated spectral function is
$\lambda = 0.438$, which
is in good agreement with other linear-response \cite{Savrasov94}
and frozen-phonon calculations \cite{Dacorogna85},
and with heat capacity data \cite{lambda_Al}.
Within Eliashberg theory, $T_c$ is a functional of the spectral
function $\alpha^2F(\omega)$ and the Coulomb parameter $\mu^*$.
Using the calculated $\alpha^2F$ as input into the
Eliashberg equations,
we find that a $\mu^*$ of 0.162 is needed to obtain the measured
transition temperature of $T_c$=1.18 K.  The same value of
$\mu^*$ yields a gap equal to the experimental value of 0.180 meV.
This consistency between independent fits to the gap and
to $T_c$ is a measure of the accuracy of our results for $\alpha^2F$.

We consider next the case of the strong-coupling superconductor Pb,
for which high-quality conventional tunneling data are
available. From the theoretical stand point, the importance
of relativistic effects in Pb make it a more difficult
system to treat than Al.
The present calculations for Pb are performed in the scalar
relativistic approximation.
The phonon dispersion curves are shown in Fig. 2.  Overall, there
is good agreement between calculated (solid lines) and
measured \cite{Pb_neutrons} (circles) phonon
frequencies, and the calculation is able to reproduce
some subtle features in the dispersion curves such as the Kohn
anomaly in the longitudinal mode along the $\Gamma$ to K direction.
Note however that  there are significant quantitative discrepancies
between the calculated and measured frequencies for the
low-energy transverse mode, especially in the regions
near X and K.
The minima in both the longitudinal  and transverse modes
at X are not observed in other fcc metals, and they
suggest the presence of very-long-range forces.  Indeed, as shown
by the dashed curves in Fig. 2, an eighth-neighbor
Born-von Karman fit \cite{Cowley}  to
the measured frequencies is unable to reproduce the
dispersion near X, especially in the case of the longitudinal mode.
Over the years, a number of exotic mechanisms have been
proposed to explain the unusual shape of the dispersion curves
in Pb \cite{Pb_phonons_theory}.
Our preliminary results obtained using the frozen-phonon approach
indicate that  differences between the calculated and
measured frequencies are significantly reduced
if the spin-orbit interaction is taken into account.

The Eliashberg function for Pb is plotted in
Fig. 3.    The calculations (solid line) yield the two-peak
structure seen in the data (circles) \cite{Pb_a2F},
but there are differences in the peak locations and
heights, especially in the case of the lower-frequency peak.
The mass enhancement parameter is calculated to be
$\lambda = 1.20$, which is significantly lower than the
tunneling result of 1.55.  These discrepancies are due in part
to the errors in the calculated transverse-mode phonon frequencies.
Note that the location of peaks in the spectral function
is determined to a large extent by the location of peaks
in the phonon density of states $F(\omega)$ ({\it i.e.,} by the delta
functions in frequency in Eq. (3)).
The overestimation of the transverse-mode
frequencies in our calculation results in an upward shift of
the lower-frequency peak in both $F(\omega)$ and $\alpha^2F(\omega)$.
We have also computed the spectral function using
the phonon frequencies generated from the Born-von Karman
fit to the data along with the calculated phonon
linewidths \cite{footnote}.
The resulting $\alpha^2F(\omega)$ is plotted as a dashed
line in Fig. 3.
Using the empirical force constants, which accurately describe the
dispersion of the transverse modes,
we obtain good agreement with the experimental results
in the low-frequency regime.
On the other hand,  since the Born-von Karman fit does
not yield accurate frequencies for the  longitudinal mode,
the resulting spectral function is less accurate  than
the first-principles result  in the high-frequency regime.
It appears that
in order to improve our description of the electron-phonon
coupling in Pb, it will be necessary to modify the computational
method to take into account the relativistic
spin-orbit coupling interaction.

Finally, we examine the electron-phonon coupling in bcc Li.
The lack of a superconducting transition in Li has been a
long-standing puzzle.
Both frozen-phonon \cite{Liu91} and RMT \cite{Li_rmt}
calculations have suggested that the electron-phonon
coupling strength in Li is similar to that in Al.
This would suggest a transition temperature on the
order of 1 K if a value of $\mu^* \approx 0.15$ is
assumed. Experimentally, however, no transition is observed, at
least down to 6 mK \cite{Li_expt}.

The LDA tends to underestimate the lattice constant
of alkali metals.  For Li, the calculated lattice constant
of 3.41 \AA~ is about 2.3\% smaller than the measured value
of 3.49 \AA.  We have performed calculations at both values of the
lattice constant.   The phonon frequencies calculated at the
experimental lattice constant are plotted in Fig. 4.  Overall,
the frequencies agree well with the neutron diffraction
data \cite{Smith68}, and subtle features such as the crossing of the
longitudinal and transverse modes along the $\Gamma$ to H direction
are reproduced by our calculations.  At the LDA lattice constant,
the calculated phonon frequencies increase by up to 8\%.
This is consistent with recent supercell
calculations  carried out at the LDA lattice constant
determined without including core
corrections (3.35 \AA) \cite{Frank95}. In that
case,  a constant scale factor of about 0.86 was needed to
bring the theoretical results in line with the
experimental frequencies.

The mass enhancement
parameter in Li  is calculated to be  $\lambda$ = 0.45 and 0.51
at the experimental and (core-corrected) LDA lattice constants,
respectively. These values are similar to results
obtained earlier within the RMT approximation  or
using the frozen-phonon method.
Using our result for $\alpha^2F(\omega)$ as
input into the Eliashberg equations, we have calculated
the superconducting transition temperature as a function of the
Coulomb pseudopotential. As shown in Fig. 5,
an unphysically large value of $\mu^* \approx 0.28$  is required
in order to suppress $T_c$ below the experimental limit.
The observed absence of superconductivity in Li therefore
remains an open problem. The resolution of this puzzle may
require consideration of the low-temperature crystal structure of Li,
in which the electron-phonon interaction may be weaker \cite{Liu91},
as well as the role of manybody interactions such
as electronic correlations and spin fluctuations \cite{Jarlborg88}.

In summary, we have presented an accurate and  efficient
method for calculating electron-phonon coupling parameters
from first principles.
The electronic response to the atomic displacements
is determined self-consistently and phonons of arbitrary wavevector
can be treated.   This method, which is an extension of the
plane-wave-based linear-response method for
calculating lattice dynamics,  is applicable
to a wide range of materials.
Results for the mass enhancement
parameter $\lambda$ and the electron-phonon
spectral function $\alpha^2F(\omega)$ for Al are in excellent
agreement with available experimental data.
In the case of Pb, there are larger
discrepancies between  our results and the tunneling data,
but we attribute this to the neglect of the spin-orbit
interaction in our calculations.  The present
results for the mass enhancement factor in bcc Li are in
accord with earlier calculations, indicating that
further theoretical investigations are needed to
resolve the question of the absence of a  superconducting
transition in this material.

\acknowledgments

We thank J. K. Freericks  and E. Nicol  for discussions
on the strong-coupling theory of superconductivity and
for their expertise in solving the Eliashberg gap equations.
AYL acknowledges the support of the Clare Boothe Luce Fund.
This work was supported in part by
the United States Department of Energy, Office
of Basic Science, Division of Materials Science.

\begin{figure}
\caption{Electron-phonon spectral function for Al.
Results from the present calculation are represented by the
solid line. Results from proximity electron tunneling
spectroscopy experiments are indicated by the
long- and short-dashed lines.
\label{Al}}
\end{figure}

\begin{figure}
\caption{Phonon dispersion curves for Pb.
The solid lines connect frequencies calculated at the sampled
wavevectors, and the circles indicate the experimentally
measured frequencies.
The dashed line is an eighth-neighbor  Born-von Karman fit  to the
measured frequencies.
\label{Pb1}}
\end{figure}

\begin{figure}
\caption{Electron-phonon spectral function for Pb.  The results
based on the calculated frequencies are
shown as a solid line, those based on the frequencies obtained
from the force-constant fit are given by the dashed line, and the
results from the inversion of tunneling data are plotted as circles.
\label{Pb2}}
\end{figure}

\begin{figure}
\caption{Phonon dispersion curves for Li.
The solid lines connect frequencies calculated at the sampled
wavevectors, and the circles denote the experimentally measured
frequencies.
\label{Li1}}
\end{figure}

\begin{figure}
\caption{Transition temperature vs. Coulomb pseudopotential for Li.
The transition temperature, which is plotted on a $\log$ scale,  is
computed as a function of $\mu^*$  using the
calculated $\alpha^2F$ as input to the Eliashberg equations.
\label{Li2}}
\end{figure}


\begin{references}
\bibitem{footnote} As can be seen from Eqs. (1) and (2),
the phonon linewidths do not depend explicitly on the phonon
frequencies.
\end{references}
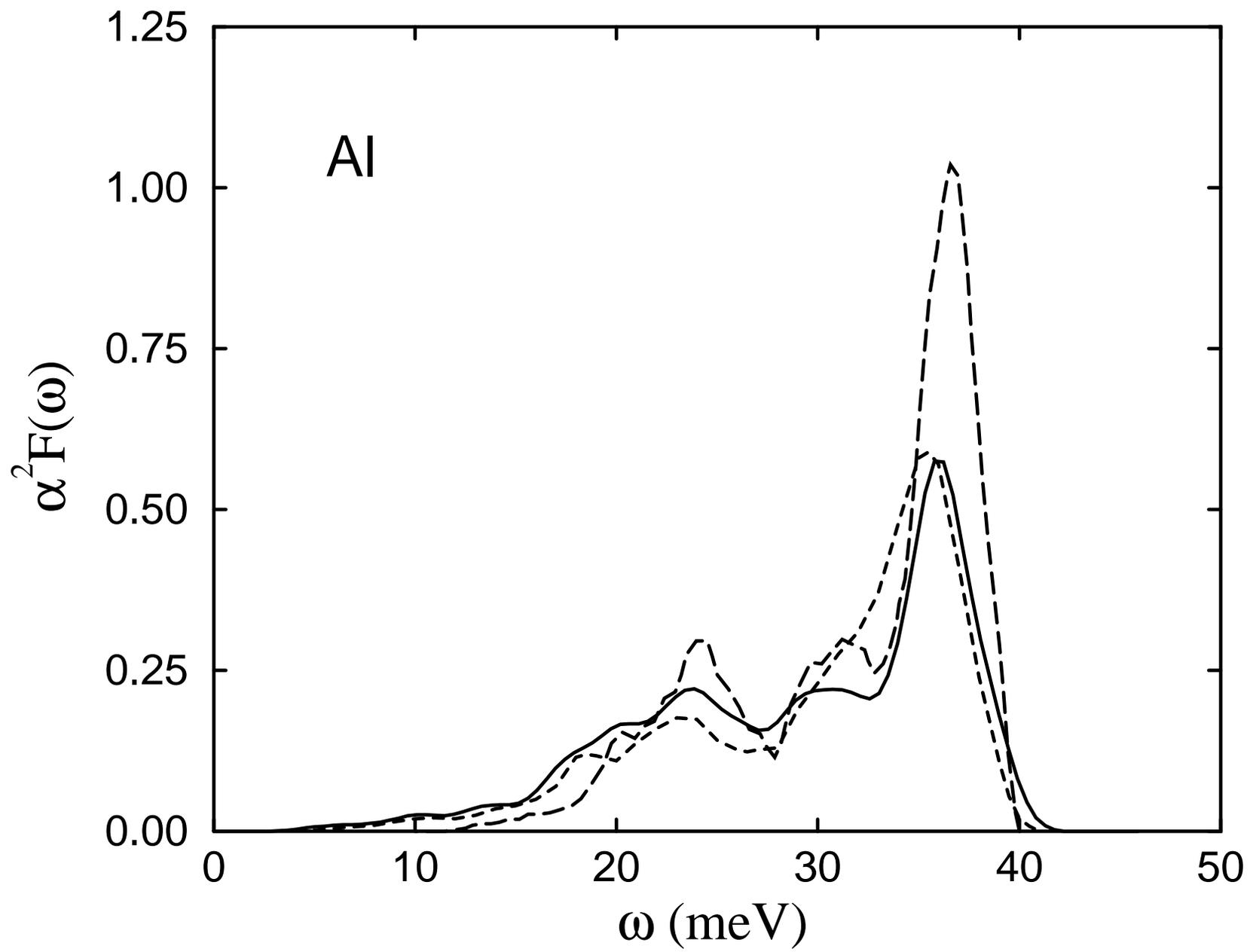
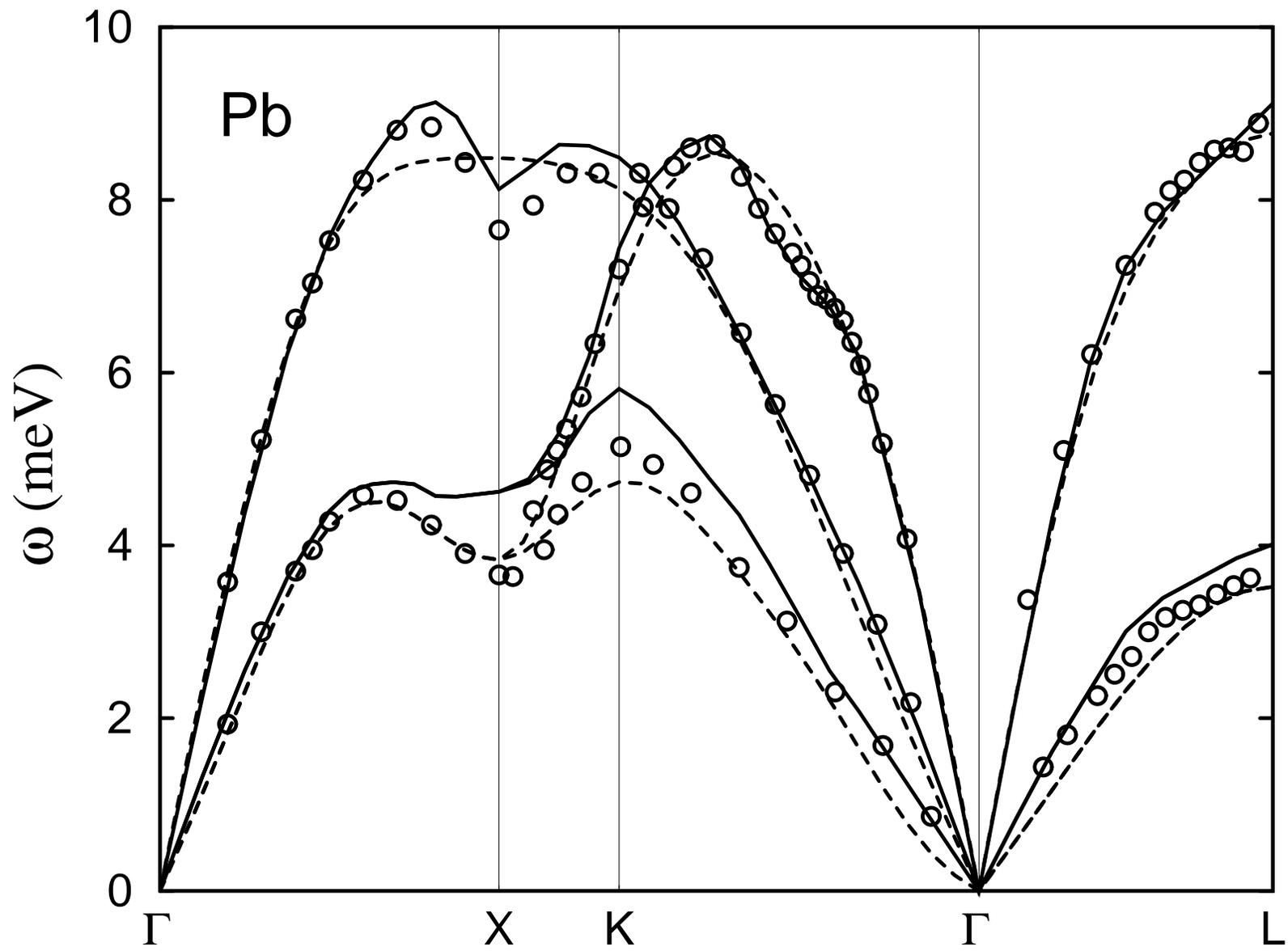
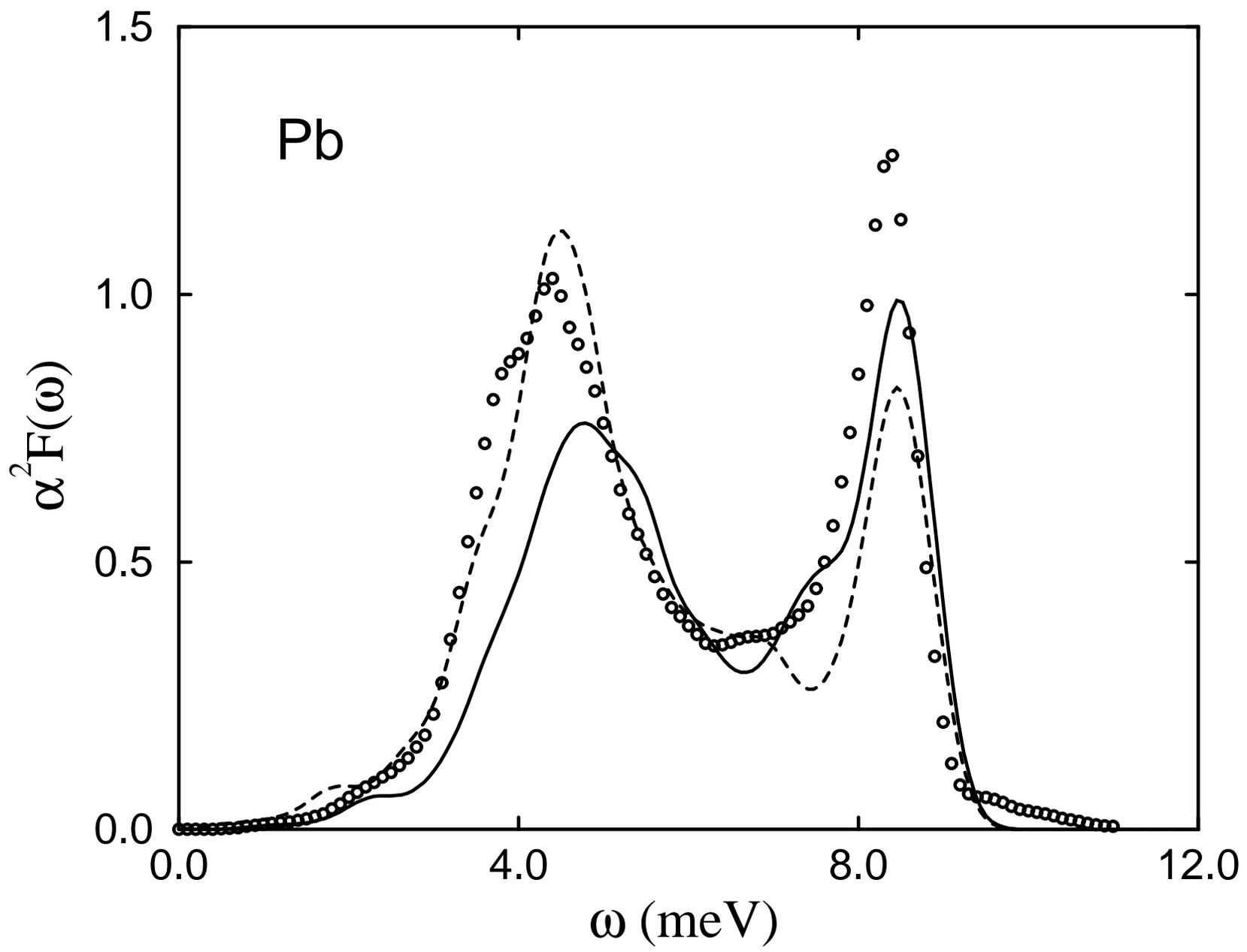
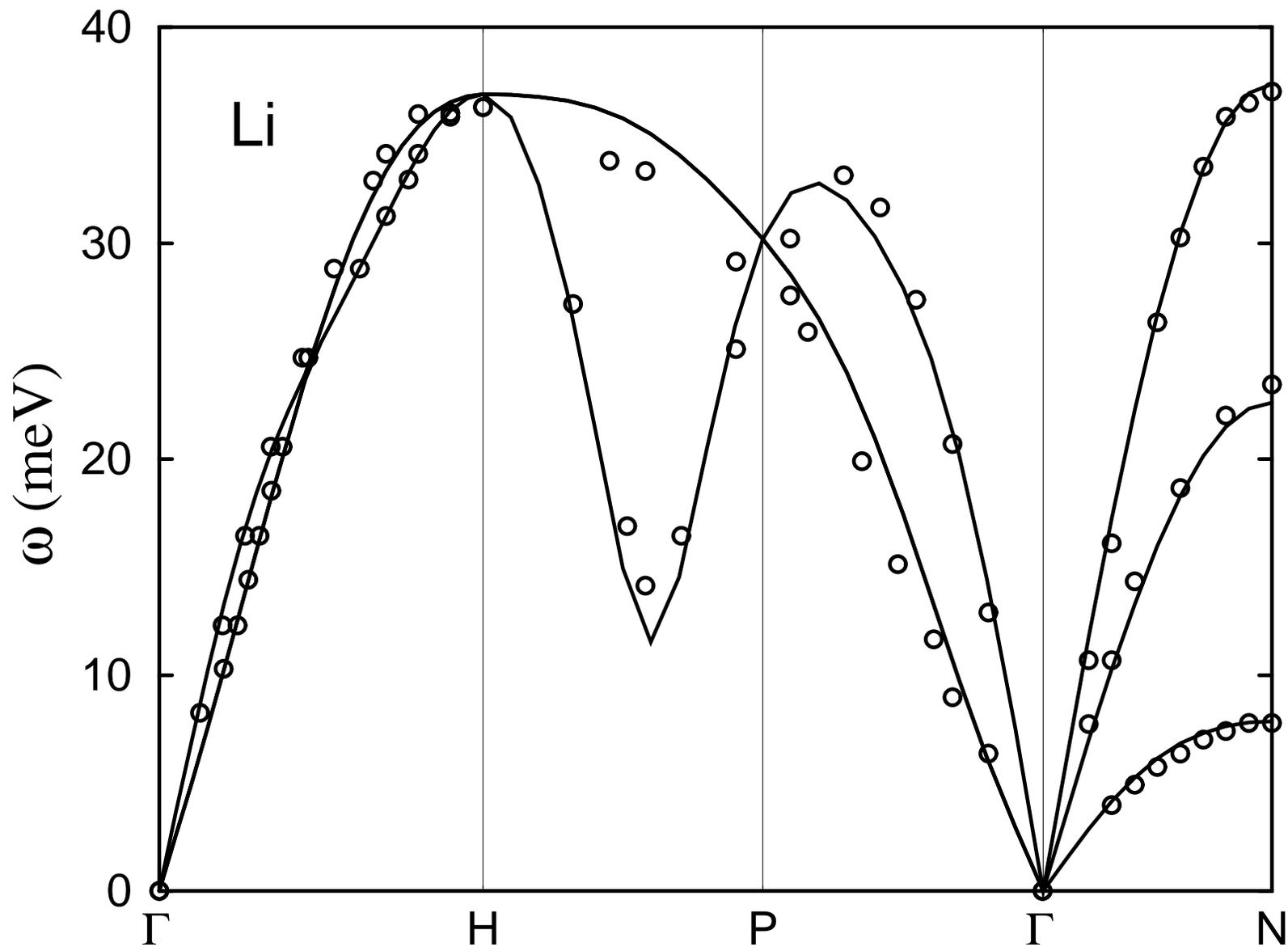
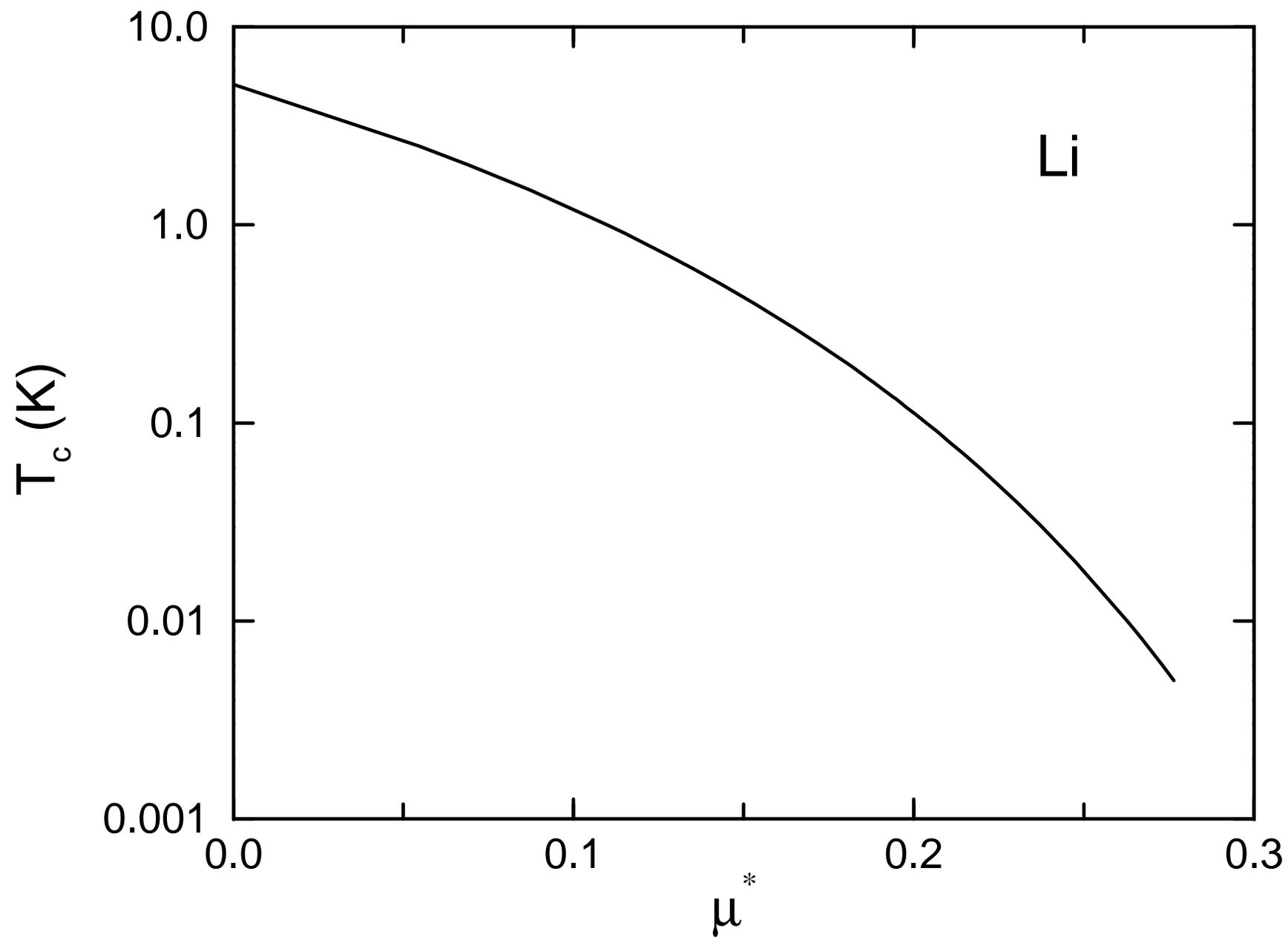
\end{document}